\documentclass{article}
\usepackage{graphicx}
\usepackage{amsmath}

\begin{document}

\title{The classical pion field in a nucleus. }
\author{Georges Ripka \and ECT*, Villa Tambosi, I-38050 Villazano (Trento), ITALY
\and Service de Physique Theorique, Centre d'Etudes de Saclay \and F-91191
Gif-sur-Yvette Cedex, FRANCE \and ECT*-07-20 preprint}
\maketitle

\begin{abstract}
A self-consistent symmetry arises when the nucleon angular momentum $j$ and
the isospin $t$ are coupled to a grand spin $G$. Closed $G$ shells become
sources of a classical pion field with a hedgehog shape. Although the
amplitude of the pion field, as measured by the chiral angle, is small, it
is found to perturb significantly the energies of the nucleon orbits.
\end{abstract}

\section{Introduction.}

\setcounter{equation}{0}\renewcommand{\theequation}{\arabic{section}.%
\arabic{equation}}

The purpose of this paper is to explore some nucleon configurations which
give rise to a classical pion field. In most mean field calculations, the
symmetries assumed for the intrinsic state do not sustain a classical
pseudoscalar pion field. For example, a closed $j$ shell can only give rise
to a spherically symmetric mean field.\ The pseudoscalar classical pion
field is expected to be composed of odd-$l$ spherical harmonics so that a
closed $j$ shell does not provide a source for the pion field.\ Nor do the
usually assumed deformed intrinsic states of even-even nuclei. This is why
only the exchange (Fock) term of the pion exchange interaction contributes
to the mean field in these nuclei, while the direct (Hartree) term vanishes.

This is a feature of mean field calculations in which the nucleus is
described by a Slater determinant or a quasi-particle vacuum. However, it
has been known for decades \cite{Brown67} that a considerable fraction of
nuclear binding is due to 2-particle 2-hole excitations caused by the pion
exchange interaction. Furthermore, Monte-Carlo calculations of light nuclei
suggest that roughly 75\% of the nuclear binding is due to the pion exchange
interaction \cite{Wiringa2001}. Mean field calculations, which use Skyrme or
Gogny forces (see the review \cite{Heenen2003} and references therein)
simulate the 2-particle 2-hole contribution of the pion exchange interaction
by an effective density dependent interaction.\ Relativistic mean field
calculations (see the review \cite{Vretenar2005}) simulate the same effect
by the presence of a sigma meson. However, the effective interactions (or
mesons) used in such mean field calculations are not able to explore the
possibility that the pion field may favor somewhat different nucleon
configurations which give rise to a finite classical pion field.

Such a possibility has however been studied by Sugimoto, Ikeda, Toki \textit{%
et al}.\cite{Toki2002,Toki2004-1,Toki2006,Toki2007}. They construct an
independent particle state composed of either mixed parity nucleon orbits or
mixed parity and isospin orbits. The charge and parity of this state are
then projected and the energy of the projected state is minimized. They find
that the tensor force, which is part of the pion exchange interaction, plays
an important role, which is understandable insofar as the projection adds
2-particle 2-hole admixtures to the Hartree-Fock state.

In this work, we investigate a self-consistent symmetry of the pion field
which leads us to construct closed $G$-shells where $\vec{G}=$ $\vec{j}+\vec{%
t}$ is the sum of the angular momentum and isospin of a nucleon orbit.

\section{The non relativistic limit of Weinberg nucleons interacting with a
chiral field.}

\setcounter{equation}{0}\renewcommand{\theequation}{\arabic{section}.%
\arabic{equation}}

The lagrangian of Weinberg nucleons interacting with a chiral field $U=\exp
\left( i\gamma _{5}\theta _{a}\tau _{a}\right) $ is\cite{Weinberg1996}: 
\begin{equation}
L=\bar{\psi}\left( i\partial _{\mu }\gamma ^{\mu }-M+i\gamma ^{\mu }U^{\frac{%
1}{2}}\left( \partial _{\mu }U^{-\frac{1}{2}}\right) \right) \psi +\frac{%
f_{\pi }^{2}}{4\nu }tr\left[ \left( \partial _{\mu }U\right) \left( \partial
^{\mu }U^{\dagger }\right) \right] +\frac{f_{\pi }^{2}m_{\pi }^{2}}{4\nu }%
tr\left( U+U^{\dagger }\right)   \label{lphiubis}
\end{equation}
It can be written in terms of the chiral angle $\theta _{a}$ as follows: 
\begin{equation*}
L=\bar{\psi}\left( i\partial _{\mu }\gamma ^{\mu }-M\right) \psi 
\end{equation*}
\begin{equation*}
+\bar{\psi}\left( -\varepsilon _{abc}\tau _{a}\sin ^{2}\frac{\theta }{2}%
\frac{\theta _{b}}{\theta }\left( \gamma ^{\mu }\partial _{\mu }\frac{\theta
_{c}}{\theta }\right) -g_{A}\gamma _{5}\frac{1}{2}\sin \theta \left( \gamma
^{\mu }\partial _{\mu }\frac{\theta _{a}\tau _{a}}{\theta }\right)
-g_{A}\gamma _{5}\frac{1}{2}\frac{\theta _{a}\tau _{a}}{\theta }\left(
\gamma ^{\mu }\partial _{\mu }\theta \right) \right) \psi 
\end{equation*}
\begin{equation}
+\frac{1}{2}f_{\pi }^{2}\left( \left( \partial _{\mu }\theta \right) \left(
\partial ^{\mu }\theta \right) +\sin ^{2}\theta \left( \partial _{\mu }\frac{%
\theta _{a}}{\theta }\right) \left( \partial ^{\mu }\frac{\theta _{a}}{%
\theta }\right) -4m_{\pi }^{2}\sin ^{2}\frac{\theta }{2}\right) 
\label{nllwphiu}
\end{equation}
where $\psi $ is the nucleon field, $\theta ^{2}=\sum_{a=1,3}\theta _{a}^{2}$
and where the last two terms in the second line have been multiplied by the
axial coupling constant $g_{A}$\cite{Weinberg1996}. The pion decay constant $%
f_{\pi }$, the axial coupling constant $g_{A}$, the pion and nucleon masses $%
m_{\pi }$ and $M$ are respectively: 
\begin{equation}
f_{\pi }=93.2\;MeV\;\;\;g_{A}=1.257\;\;\;m_{\pi
}=139.6\;MeV\;\;\;M=938.9\;MeV
\end{equation}
Our results (see Fig.\ref{fig:rhotet}) yield chiral angles $\theta _{a}$
which do not exceed $0.05$ so that we can safely work with the low amplitude
limit, in which the chiral field reduces to a pion field.\ For small values
of $\theta _{a}$, the lagrangian (\ref{nllwphiu}) reduces to: 
\begin{equation}
L=\bar{\psi}\left( i\partial _{\mu }\gamma ^{\mu }-M-\frac{1}{2}g_{A}\gamma
_{5}\gamma ^{\mu }\tau _{a}\left( \partial _{\mu }\theta _{a}\right) \right)
\psi +\frac{1}{2}f_{\pi }^{2}\left( \left( \partial _{\mu }\theta
_{a}\right) \left( \partial ^{\mu }\theta _{a}\right) -m_{\pi }^{2}\theta
_{a}^{2}\right) 
\end{equation}
and the pion field $\pi _{a}$ is related to the chiral angle $\theta _{a}$
by: 
\begin{equation}
\pi _{a}=f_{\pi }\theta _{a}
\end{equation}
The mean field approximation consists in assuming that the chiral (or pion)
field is classical.\ We shall further consider time-independent stationary
states, in which case we can work with the hamiltonian: 
\begin{equation}
H=\int d^{3}x\;\psi ^{\dagger }\left( \frac{\vec{\alpha}\cdot \vec{\nabla}}{i%
}+\beta M-\frac{1}{2}g_{A}\gamma _{5}\tau _{a}\left( \vec{\alpha}\cdot \vec{%
\nabla}\theta _{a}\right) \right) \psi +\frac{1}{2}f_{\pi }^{2}\int
d^{3}x\left( \left( \vec{\nabla}\theta _{a}\right) ^{2}+m_{\pi }^{2}\theta
_{a}^{2}\right) 
\end{equation}

We further proceed to make a non relativistic reduction of the Dirac
hamiltonian: 
\begin{equation}
h=\frac{\vec{\alpha}\cdot\vec{\nabla}}{i}+\beta M-\frac{1}{2}g_{A}\gamma
_{5}\tau_{a}\left( \vec{\alpha}\cdot\vec{\nabla}\theta_{a}\right)
\end{equation}
The reduction is made in two steps.\ First we maintain the nucleon orbits
orthogonal to the vacuum Dirac sea orbits by expanding the eigenstates of $h$
on the \emph{positive energy} nucleon plane wave states $\left| \vec
{k},\sigma,\tau,+\right\rangle $, which are eigenstates of $\frac{\vec{%
\alpha }\cdot\vec{\nabla}}{i}+\beta M$ belonging to the positive eigenvalues 
$\sqrt{k^{2}+M^{2}}$.\ This is achieved by working with a modified Dirac
hamiltonian $h_{NR}$ defined such that it has the same matrix elements as $h$
between the positive energy plane wave states: 
\begin{equation}
\left\langle \vec{k},\sigma,\tau\left| h_{NR}\right| \vec{k}%
^{\prime},\sigma^{\prime},\tau^{\prime}\right\rangle \equiv\left\langle \vec{%
k},\sigma,\tau,+\left| h\right| \vec{k}^{\prime},\sigma^{\prime},\tau^{%
\prime },+\right\rangle
\end{equation}
The second step consists in expanding the resulting $h_{NR}$ in powers of $%
k/M$. When the expansion is limited to order $k^{2}/M^{2}$ the resulting non
relativistic hamiltonian acquires the familiar form: 
\begin{equation}
h_{NR}=-\frac{\nabla^{2}}{2M}-\frac{g_{A}}{2}\sigma_{i}\tau_{a}\left(
\nabla_{i}\theta_{a}\right)
\end{equation}

\section{The model hamiltonian and $G$ shells.}

\setcounter{equation}{0}\renewcommand{\theequation}{\arabic{section}.%
\arabic{equation}}

In order to explore nucleon configurations which are likely to give rise to
a classical pion field, we further simplify the dynamics by assuming that
the nucleons, which interact with the classical pion field, are in spherical
oscillator orbits and that they also interact with a spin-orbit contact
interaction of the form\cite{Heenen2003}: 
\begin{equation*}
V_{ls}=iW_{ls}\frac{1}{2}\int d^{3}r_{1}d^{3}r_{2}\;
\end{equation*}
\begin{equation}
\psi ^{\dagger }\left( \vec{r}_{1}\right) \psi ^{\dagger }\left( \vec{r}%
_{2}\right) \left[ \varepsilon _{ijk}\left( \sigma _{1}^{i}+\sigma
_{2}^{i}\right) \left( p_{1}^{j}-p_{2}^{j}\right) \delta \left( \vec{r}_{1}-%
\vec{r}_{2}\right) \left( p_{1}^{k}-p_{2}^{k}\right) \right] \psi \left( 
\vec{r}_{2}\right) \psi \left( \vec{r}_{1}\right)  \label{vls}
\end{equation}
where $\vec{p}=\vec{\nabla}/i$.\ The expectation value of the spin-orbit
interaction $V_{ls}$ in a Slater determinant $\left| \Phi \right\rangle $
can be reduced to the form: 
\begin{equation}
\left\langle \Phi \left| V_{ls}\right| \Phi \right\rangle =-4W_{ls}\int
d^{3}r\;\vec{B}\left( \vec{r}\right) \cdot \left( \vec{\nabla}n\left( \vec{r}%
\right) \right)  \label{vlsred}
\end{equation}
where $n\left( \vec{r}\right) $ and $\vec{B}\left( \vec{r}\right) $ are the
following nucleon densities: 
\begin{equation}
n\left( \vec{r}\right) =\left\langle \Phi \left| \psi ^{\dagger }\left( \vec{%
r}\right) \psi \left( \vec{r}\right) \right| \Phi \right\rangle \;\;\;\;\;\;%
\vec{B}\left( \vec{r}\right) =\left\langle \Phi \left| \psi ^{\dagger
}\left( \vec{r}\right) \left( \vec{\sigma}\times \vec{p}\right) \psi \left( 
\vec{r}\right) \right| \Phi \right\rangle  \label{nbden}
\end{equation}

The model hamiltonian is: 
\begin{equation}
H=\int d^{3}x\;\psi^{\dagger}\left( h_{osc}-\frac{g_{A}}{2}\sigma_{i}\tau
_{a}\left( \nabla_{i}\theta_{a}\right) \right) \psi+V_{ls}+\frac{1}{2}%
f_{\pi}^{2}\int d^{3}x\left( \left( \vec{\nabla}\theta_{a}\right)
^{2}+m_{\pi}^{2}\theta_{a}^{2}\right)  \label{ham1}
\end{equation}
where 
\begin{equation}
h_{osc}=-\frac{\nabla^{2}}{2M}+\frac{1}{2}M\omega^{2}r^{2}
\end{equation}

We shall diagonalize the single particle hamiltonian in a basis in which the
angular momentum $\vec{j}$ and the isospin $\vec{t}$ are coupled to a grand
spin $G$: 
\begin{equation}
\left| nljGM\right\rangle =\sum_{mm_{t}}\left\langle jm,\frac{1}{2}%
m_{t}\left| GM\right. \right\rangle \left| nljm\right\rangle \left| \frac{1}{%
2}m_{t}\right\rangle  \label{ljgm}
\end{equation}
where $\left| nljm\right\rangle $ is a usual spherical shell model orbit and 
$\left\langle jm,\frac{1}{2}m_{t}\left| GM\right. \right\rangle $ a
Clebsch-Gordan coefficient. In this basis, the single particle spin-orbit
interaction remains diagonal.\ We will show that there is a self-consistent
symmetry in which nucleons form closed $G$-shells and the pion field has a
hedgehog shape. To any non-zero grand spin and projection $\left( G,M\right) 
$ there correspond four states, which we denote as $\left| GMi\right\rangle $
with $i=1,2,3,4$, namely: 
\begin{equation}
\begin{tabular}{|l|l|l|l|}
\hline
$\left| GM1\right\rangle $ & $l=G-1=j-\frac{1}{2}$ & $j=l+\frac{1}{2}=G-%
\frac{1}{2}$ & $P=\left( -1\right) ^{G+1}$ \\ \hline
$\left| GM2\right\rangle $ & $l=G+1=j+\frac{1}{2}$ & $j=l-\frac{1}{2}=G+%
\frac{1}{2}$ & $P=\left( -1\right) ^{G+1}$ \\ \hline
$\left| GM3\right\rangle $ & $l=G=j-\frac{1}{2}$ & $j=l+\frac{1}{2}=G+\frac{1%
}{2}$ & $P=\left( -1\right) ^{G}$ \\ \hline
$\left| GM4\right\rangle $ & $l=G=j+\frac{1}{2}$ & $j=l-\frac{1}{2}=G-\frac{1%
}{2.}$ & $P=\left( -1\right) ^{G}$ \\ \hline
\end{tabular}
\label{jgi}
\end{equation}
To grand spin $G=0$ states correspond two shell model states: 
\begin{equation}
\begin{tabular}{|l|l|l|l|}
\hline
$\left| \left( G=0\right) ,1\right\rangle $ & $l=G=0$ & $j=l+\frac{1}{2}=G+%
\frac{1}{2}=\frac{1}{2}$ & $P=+$ \\ \hline
$\left| \left( G=0\right) ,2\right\rangle $ & $l=G+1=1$ & $j=l-\frac{1}{2}=G+%
\frac{1}{2}=\frac{1}{2}$ & $P=-$ \\ \hline
\end{tabular}
\label{jgizero}
\end{equation}
For example, the four $\left| nGMi\right\rangle $ states with $n=1$ and $G=2$
are: 
\begin{equation}
\begin{tabular}{|l|l|l|l|l|}
\hline
$G$ & $i$ & $l$ & $j$ &  \\ \hline
$2$ & $1$ & $1$ & $\frac{3}{2}$ & $1p_{\frac{3}{2}}$ \\ \hline
$2$ & $2$ & $3$ & $\frac{5}{2}$ & $1f_{\frac{5}{2}}$ \\ \hline
$2$ & $3$ & $2$ & $\frac{5}{2}$ & $1d_{\frac{5}{2}}$ \\ \hline
$2$ & $4$ & $2$ & $\frac{3}{2}$ & $1d_{\frac{3}{2}}$ \\ \hline
\end{tabular}
\end{equation}
and the two states with $n=1$ and $G=0$ are the $1s_{\frac{1}{2}}$ and $1p_{%
\frac{1}{2}}$ states.

\section{The self-consistent symmetry generated by closed $G$ shells.}

\setcounter{equation}{0}\renewcommand{\theequation}{\arabic{section}.%
\arabic{equation}}

A closed $\left( nGi\right) $ shell is a state in which the $2G+1$ states $%
\left| nGMi\right\rangle $ with different values of $M$ are occupied.\
Applying Wick's theorem, we find that the expectation value of a tensor
operator of rank $K$ and component $q$ in a closed $\left( nGi\right) $
shell is 
\begin{equation}
\sum_{M}\left\langle nGMi\left| T_{Kq}\right| nGMi\right\rangle =\delta _{K0}%
\sqrt{2G+1}\left\langle nGi\left| \left| T_{K}\right| \right|
nGi\right\rangle
\end{equation}
so that it vanishes unless $K=0$. When applying Wick's theorem, one should
remember that the rank $K$ of the tensor refers to rotations in the combined 
$\left( \vec{r},\sigma,\tau\right) $ space which includes rotations in
isospin space. The expectation value, in a closed $G$ shell, of the
classical pion field appearing in the hamiltonian (\ref{ham1}) is: 
\begin{equation}
\mathcal{E}=-\frac{g_{A}}{2}\sum_{M}\left\langle nGMi\left| \tau_{a}\vec{%
\sigma}\cdot\vec{\nabla}\theta_{a}\right| nGMi\right\rangle
\end{equation}
To prevent $\mathcal{E}$ from vanishing, $\tau_{a}\vec{\sigma}\cdot\vec
{\nabla}\theta_{a}$ must be a tensor of rank zero in $\left( \vec{r}%
\sigma\tau\right) $ space. Since $\tau_{a}$ is a tensor of rank $1$ in this
space, the condition for $\mathcal{E}$ not to vanish is that $\vec{\sigma }%
\cdot\vec{\nabla}\theta_{a}$ should also be a tensor of rank 1. However, $%
\vec{\sigma}\cdot\vec{\nabla}$ is a scalar so that the condition requires $%
\theta_{a}$ to be a vector and this in turn means that the chiral angle must
have the so-called \emph{hedgehog shape}: 
\begin{equation}
\theta_{a}\left( \vec{r}\right) =\widehat{r}_{a}\theta\left( r\right)
\label{hedgehog}
\end{equation}
where $\widehat{r}=\vec{r}/r$. This also implies that closed $G$-shells have
definite parity.

If we calculate $\left( \nabla_{i}\theta_{a}\right) $ with the hedgehog
field (\ref{hedgehog}), we find: 
\begin{equation}
\sigma_{i}\tau_{a}\left( \nabla_{i}\widehat{r}_{a}\theta\right) =\vec
{\sigma}\cdot\vec{\tau}\frac{\theta}{r}+\left( \widehat{r}\cdot\vec{\sigma }%
\right) \left( \widehat{r}\cdot\vec{\tau}\right) \left( \frac{d\theta }{dr}-%
\frac{\theta}{r}\right)  \label{sigtau1}
\end{equation}
A lengthy although straightforward calculation will also reveal that the
two-body spin orbit interaction $V_{ls}$ gives rise to a single-particle
potential of the form: 
\begin{equation}
h_{ls}=4W_{ls}\frac{1}{r}\frac{dn}{dr}\vec{\sigma}\cdot\vec{L}+4W_{ls}\left( 
\frac{1}{r}\frac{dB}{dr}+\frac{B}{r^{2}}\right)  \label{hls1}
\end{equation}
where $\vec{L}=\vec{r}\times\vec{p}$, and where $n\left( r\right) $ and $%
B\left( r\right) $ are related to the densities (\ref{nbden}) as follows: 
\begin{equation}
n\left( \vec{r}\right) =n\left( r\right) \;\;\;\;\vec{B}\left( \vec
{r}\right) =\widehat{r}B\left( r\right)
\end{equation}
Each closed $\left( nGi\right) $ shell makes the following contribution to
these densities: 
\begin{equation}
n_{nGi}\left( r\right) =\sum_{M}\left\langle nGMi\left| \vec{r}\right.
\right\rangle \left\langle \vec{r}\left| nGMi\right. \right\rangle
\;\;\;\;B_{nGi}\left( r\right) =-\sum_{M}\left\langle nGMi\left| \vec
{r}\right. \right\rangle \frac{1}{r}\vec{\sigma}\cdot\vec{L}\left\langle 
\vec{r}\left| nGMi\right. \right\rangle
\end{equation}
The total densities $n\left( r\right) $ and $B\left( r\right) $ are obtained
by summing these contributions over the \emph{occupied} $\left( nGi\right) $
shells. Thus the single particle hamiltonian which gives rise to closed $G$
shells is: 
\begin{equation*}
h_{NR}=-\frac{\nabla^{2}}{2M}+\frac{1}{2}M\omega^{2}r^{2}-\frac{g_{A}}{2}%
\left[ \vec{\sigma}\cdot\vec{\tau}\frac{\theta}{r}+\left( \widehat {r}\cdot%
\vec{\sigma}\right) \left( \widehat{r}\cdot\vec{\tau}\right) \left( \frac{%
d\theta}{dr}-\frac{\theta}{r}\right) \right]
\end{equation*}

\begin{equation}
+4W_{ls}\frac{1}{r}\frac{dn}{dr}\vec{\sigma}\cdot\vec{L}+4W_{ls}\left( \frac{%
1}{r}\frac{dB}{dr}+\frac{B}{r^{2}}\right)  \label{hnr}
\end{equation}

The expectation value of the third component $t_{3}=\frac{1}{2}\tau_{3}$ of
isospin in a $\left| nGMi\right\rangle $ orbit is: 
\begin{equation}
\left\langle nGMi\left| t_{3}\right| nGMi\right\rangle =\frac{M}{2G}\;\left(
i=1,4\right) \;\;\;\;\left\langle nGMi\left| t_{3}\right| nGMi\right\rangle
=-\frac{M}{2\left( G+1\right) }\;\left( i=2,3\right)
\end{equation}
$\allowbreak$It follows that a closed $G$ shell has an equal \emph{average}
numbers of neutrons and protons. Furthermore, a \emph{closed} $G$ shell
contains $2G+1$ orbits and it therefore contains an odd number of nucleons.

\section{Nucleon orbits in a hedgehog field of varying amplitude.}

\setcounter{equation}{0}\renewcommand{\theequation}{\arabic{section}.%
\arabic{equation}}

A self-consistent calculation of the chiral angle requires an initial guess
of either the nucleon orbits or of the chiral angle $\theta \left( r\right) $%
. In this respect, a useful guide consists in assuming a simple form for the
chiral angle.\ Since it is a $p$ wave, it likely to grow linearly with $r$
at the origin.\ It is also likely to decrease exponentially at infinity as $%
\exp \left( -m_{\pi }r\right) $.\ The simplest function which will do this
is: 
\begin{equation}
\theta \left( x\right) =axe^{-x}\;\;\;\;x\equiv m_{\pi }r  \label{theta}
\end{equation}
where $a$ measures the amplitude of the chiral angle. The self-consistent
calculation described in section \ref{sec:selfcon} shows that (\ref{theta})
is quite a good initial guess for the shape of the chiral angle. An example
is displayed in Fig.\ref{fig:rhotet}. We can then diagonalize the single
particle hamiltonian (\ref{hnr}) with and plot the spectrum of the nucleon
orbits as a function of the amplitude $a$ of the chiral angle (\ref{theta}).

In this exploratory calculation, we do not mix radial quantum numbers so
that each orbit has a given principal quantum number $n$.\ The nucleon
orbits $\left| \lambda\right\rangle $ are thus expanded on a $\left|
nGMi\right\rangle $ basis: 
\begin{equation}
\left| \lambda\right\rangle \equiv\left| n_{\lambda}G_{\lambda}M_{\lambda
}k_{\lambda}\right\rangle =\sum_{i}c_{i,k_{\lambda}}^{\left( n_{\lambda
}G_{\lambda}\right) }\left| n_{\lambda}G_{\lambda}M_{\lambda}i\right\rangle
\label{lngm}
\end{equation}
The sum over $i$ extends from $1$ to $4$ when $G>0$ and from $1$ to $2$ when 
$G=0$.\ However, since closed $G$ shells give rise to orbits with definite
parity only two $\left| nGMi\right\rangle $ states contribute to any one
orbit $\left| \lambda\right\rangle $.\ The coefficients $c_{i,k_{\lambda}}^{%
\left( n_{\lambda}G_{\lambda}\right) }$ are determined by diagonalizing the
hamiltonian (\ref{hnr}): 
\begin{equation}
h_{NR}\left| n_{\lambda}G_{\lambda}M_{\lambda}k_{\lambda}\right\rangle
=e_{n_{\lambda}G_{\lambda}k_{\lambda}}\left|
n_{\lambda}G_{\lambda}M_{\lambda}k_{\lambda}\right\rangle  \label{heigen}
\end{equation}
in this basis. The energy of filled $G$ shells is then the following
functional of the chiral angle $\theta$: 
\begin{equation}
E\left( \theta\right) =\sum_{n_{\lambda}G_{\lambda}k_{\lambda}\in F}\left(
2G_{\lambda}+1\right) e_{n_{\lambda}G_{\lambda}k_{\lambda}}+\frac{1}{2}%
f_{\pi}^{2}\int d^{3}r\left( \left( \frac{d\theta}{dr}\right) ^{2}+\frac
{2}{r^{2}}\theta^{2}+m_{\pi}^{2}\theta^{2}\right) +\left\langle \Phi\left|
V_{ls}\right| \Phi\right\rangle  \label{etheta}
\end{equation}
where the last term is the expectation value of the spin-orbit interaction (%
\ref{vls}). The sum is limited to occupied $G$ shells.

\begin{figure}[th]
\begin{center}
\includegraphics[width=10cm]{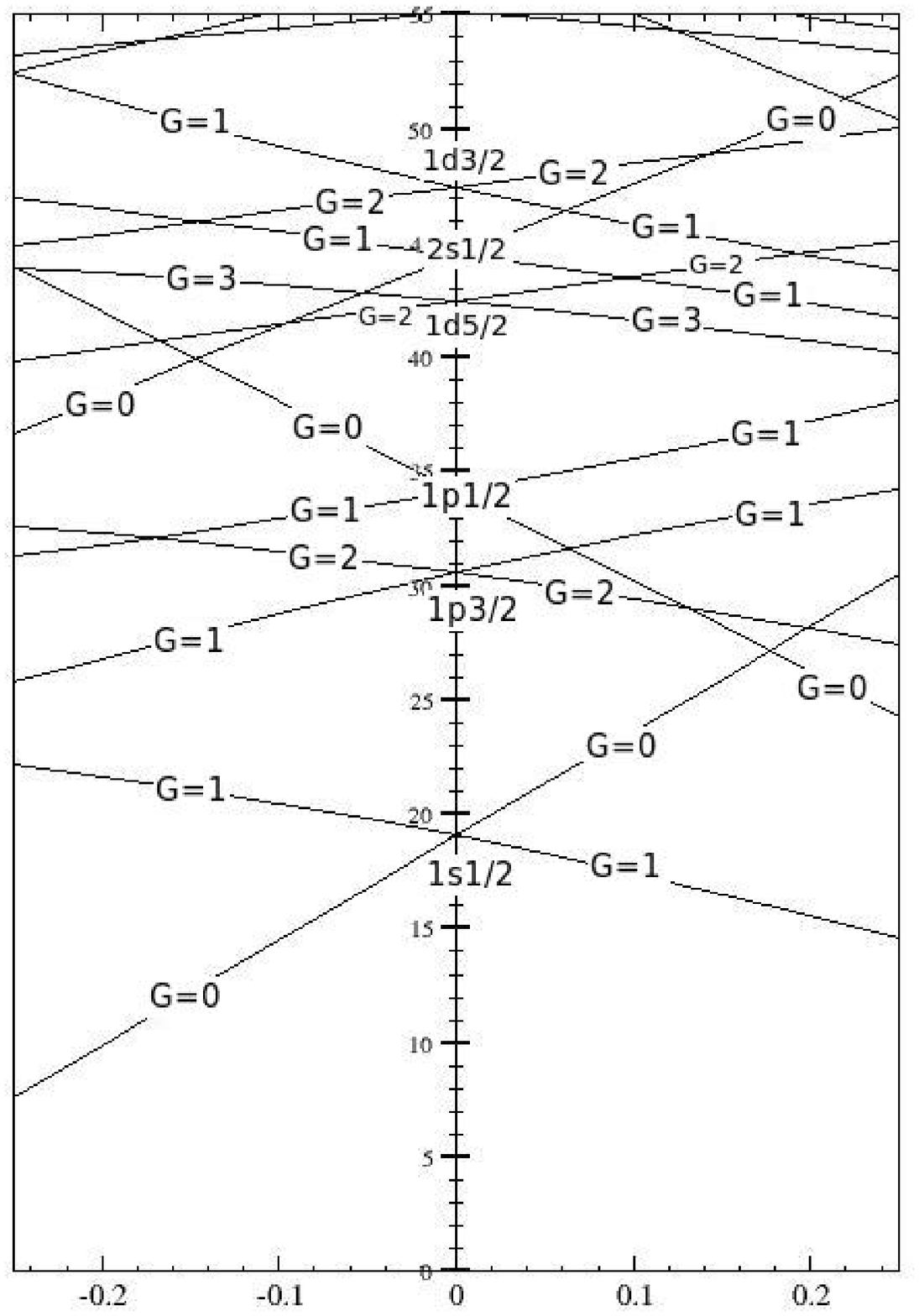}
\end{center}
\caption{The spectrum of the $\left| nGMk\right\rangle $ orbits are plotted
as a function of the amplitude $a$ of the chiral angle (\ref{theta}). When $%
a=0$ the chiral angle vanishes and the orbits reduce to spherical shell
model orbits $\left| nljm\right\rangle $. When $a\neq0$ each $j$ shell
splits into two $G$ orbits with $G=j\pm\frac{1}{2}$ as indicated on the
figure. The orbits were calculated with an oscillator frequency $\hbar%
\protect\omega=12.71\;fm$ applicable to $^{16}O$.}
\label{fig:orb16}
\end{figure}

\begin{figure}[th]
\begin{center}
\includegraphics[width=10cm]{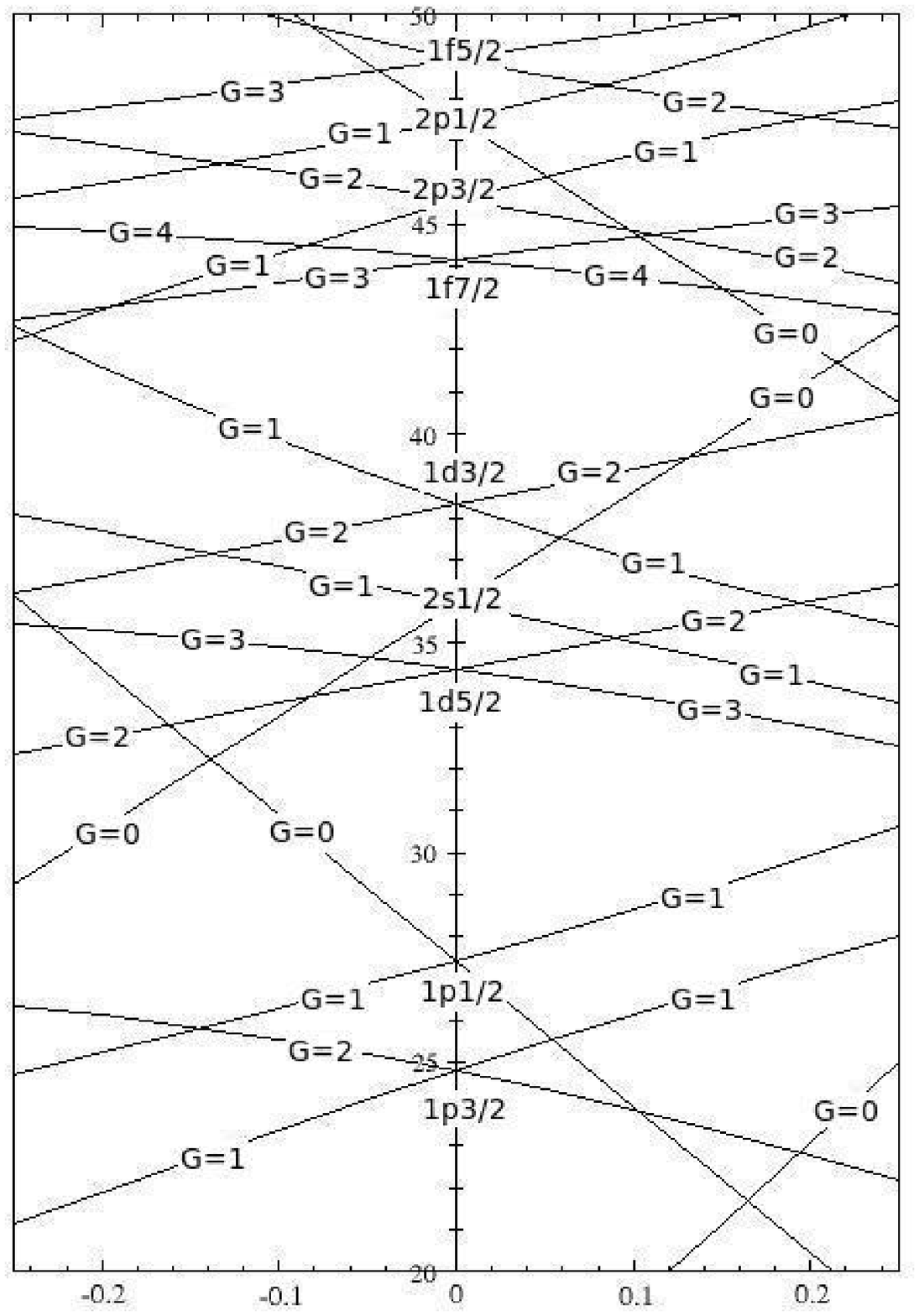}
\end{center}
\caption{The same as Fig \ref{fig:orb16} for somewhat higher energy orbits,
calculated with $\hbar\protect\omega=10.27\;MeV$ applicable to $^{40}Ca$.}
\label{fig:orbCa40}
\end{figure}

Figure \ref{fig:orb16} shows the spectrum of the low energy nucleon orbits
plotted as a function of the amplitude $a$ of the chiral angle.\ They are
calculated using a harmonic oscillator constant $\hbar \omega =12.71\;Mev$
which fits the observed charge radios $2.71\;fm$ of $^{16}O$\cite
{Hodgson1984}. Figure \ref{fig:orbCa40} shows somewhat higher energy orbits
calculated with $\hbar \omega =10.27\;MeV$ which fits the observed charge
radius $3.48\;fm$ of $^{40}Ca$. In both figures the spin-orbit interaction
strength was $W_{ls}=15\;fm^{-5}$ adjusted to yield a $5\;Mev$ splitting
between the $1d_{\frac{5}{2}}$ and $1d_{\frac{3}{2}}$ states in $^{16}O$.
When $a=0$ the chiral angle and the pion field vanish and the orbits reduce
to spherical shell model orbits $\left| nljm\right\rangle $. In the presence
of the chiral field $\left( a\neq 0\right) $ each $j$ shell of given parity
splits into two $G$ shells with $G=j\pm \frac{1}{2}$.\ 

\qquad\qquad

\section{\protect\bigskip Self consistent calculation of the chiral angle.}

\setcounter{equation}{0}\renewcommand{\theequation}{\arabic{section}.%
\arabic{equation}}

\label{sec:selfcon}

The chiral angle can be calculated by minimizing the energy (\ref{etheta}).
The equation $\delta E\left( \theta\right) /\delta\theta\left( r\right) =0 $
can be solved for $\theta\left( r\right) $ in the form: 
\begin{equation}
\theta\left( x\right) =-\frac{m_{\pi}}{2\pi^{2}f_{\pi}^{2}}\int_{0}^{\infty
}x^{\prime2}dx^{\prime}\;i_{1}\left( x_{<}\right) k_{1}\left( x_{>}\right)
\rho\left( x^{\prime}\right)  \label{tetexpl}
\end{equation}
where $x_{<}$ and $x_{>}$ are respectively the lower and upper values of $x$
and $x^{\prime}$and where the Bessel functions are: 
\begin{equation}
i_{1}\left( z\right) =-\frac{\sinh\left( z\right) }{z^{2}}+\frac
{\cosh\left( z\right) }{z}\;\;\;\;\;k_{1}\left( z\right) =\frac{\pi}{2z}%
e^{-z}\left( 1+\frac{1}{z}\right)  \label{i1k1}
\end{equation}
The source term in (\ref{tetexpl}) is calculated in terms of the occupied
orbits: 
\begin{equation}
x^{2}\rho\left( x\right) =\frac{\delta}{\delta\theta\left( x\right) }%
\sum_{n_{\lambda}G_{\lambda}k_{\lambda}\in F}\left( 2G_{\lambda}+1\right)
e_{n_{\lambda}G_{\lambda}k_{\lambda}}  \label{rhox}
\end{equation}

The spectra displayed on Figs.\ref{fig:orb16} and \ref{fig:orbCa40} can be
used to guess which filled $G$ shells are likely to give rise to a favored
configuration.\ We expect that favored configurations will be those which
give rise to an appreciable energy gap between the occupied and empty $G$%
-shells. Once we have decided which $G$ shells are filled, we can make a
self-consistent calculation of the chiral angle $\theta \left( r\right) $:

\begin{enumerate}
\item  We choose a set of closed $G$ shells and, with this set, we calculate
the source term $\rho \left( x\right) $ using (\ref{rhox}) with unperturbed
orbits.

\item  With this $\rho \left( x\right) $, we calculate $\theta \left(
x\right) $ using (\ref{tetexpl}).

\item  With this $\theta \left( x\right) $ we calculate the orbits (\ref
{lngm}) by diagonalizing the hamiltonian (\ref{hnr}).

\item  With this set of orbits we calculate $\rho \left( x\right) $ using (%
\ref{rhox}) and we return to step 2.\ We continue iterating until
convergence is reached.
\end{enumerate}

The results are summarized in Tables \ref{table1} and \ref{table2} for
configurations containing $12<A<37$ nucleons and $2p-1f$ shell
configurations respectively.\ The tables specify which closed $G$ shells are
filled, the number of nucleons, the energy gap separating filled and empty
orbits and the maximum value of the chiral angle $\theta\left( x\right) $
for each configuration. An orbit labelled, for example, $\left(
1d_{5/2}\right) \;G=2$ refers to an orbit which, on Fig.\ref{fig:orb16}, is
marked $G=2$ and which converges to the $1d_{5/2}$ state when $a=0$.
Although the self consistent orbits are not identical to those plotted on
Fig.\ref{fig:orb16}, they have a large overlap with the latter. We have only
listed cases where the filled $G$ shells have lower energies than the empty
shells.

When both $G$ shells emanating from a given $j$ shell are filled, the
resulting Slater determinant is equivalent to one in which the $j$ shell is
closed and it does not contribute to the source term $\rho \left( x\right) $%
. The closed $j$ shells are listed in the first column of tables \ref{table1}
and \ref{table2}. The second column specifies the filled and closed $G$
shells which correspond to partially filled $j$ shells. Since every $G$
shell contains an odd number of nucleons, and since a closed $G$ shell has
the same average number of neutrons and protons, configurations which have
an odd number of filled $G$ shells refer to odd-even nuclei.\ Most
configurations listed in table \ref{table1} are of this type.\ Thus, for
example, the $A=13$ configuration could refer to the mirror nuclei $%
_{7}^{13}N_{6}$ and $_{6}^{13}C_{7}$ and the $A=31$ configuration to $%
_{16}^{31}S_{15}$ and $_{15}^{31}P_{16}$.

There is a difference between the energy gaps, displayed in Figs.\ref
{fig:orb16} and \ref{fig:orbCa40}, and the energy gaps which are given in
Tables \ref{table1} and \ref{table2}. The energy gaps appearing in the
figures are plotted as a function of the strength of the pion hedgehog field
(as measured by the amplitude of the chiral angle). However, in Tables \ref
{table1} and \ref{table2}, the displayed energy gaps are obtained
self-consistently by a minimization of the energy. We see that in most
odd-even nuclei, the chiral angles which minimize the energy are
considerably smaller than the ones which produce the largest energy gaps
(displayed in Figs.\ref{fig:orb16} and \ref{fig:orbCa40}). The reason that
the energy required to produce a classical field increases with the
amplitude of the chiral angle. This is the case of most odd-even nuclei,
listed in Tables \ref{table1} and \ref{table2}, for which the chiral field
is not strong enough to produce a level crossing between filled and empty
levels.

One exception in Table \ref{table1} is the $A=22$ configuration, which has
filled $\left( 1d_{5/2}\right) \;G=2$ and $\left( 2s_{1/2}\right) \;G=0$
shells.\ Fig.\ref{fig:orb16} shows that, for $a<0$, the $\left(
2s_{1/2}\right) \;G=0$ orbit crosses the $\left( 1d_{5/2}\right) G=3$ orbit
and it does so also with the calculated chiral angle. Although it has a low
energy gap, this configuration gives rise to a relatively strong chiral
field. \ It refers to an odd-odd nucleus such as $_{11}^{22}Na_{11}$.

Another example is the $A=62$ configuration shown in Table \ref{table2}. In
this configuration, three filled $G$ shells become almost degenerate, namely
the $\left( 1f_{7/2}\right) \;G=4$, $\left( 2p_{3/2}\right) \;G=2$ and $%
\left( 2p_{1/2}\right) \;G=0$ orbits.\ This is clearly visible in Fig.\ref
{fig:orbCa40}. In the self-consistent calculation, they are also almost
degenerate, their energies being respectively $53.33,54.68$ and $54.19\;MeV$%
. This configuration also gives rise to a relatively strong chiral field. \
It refers to an odd-odd nucleus such as $_{31}^{62}Ga_{31}$.

\begin{figure}[h]
\begin{center}
\includegraphics[width=10cm]{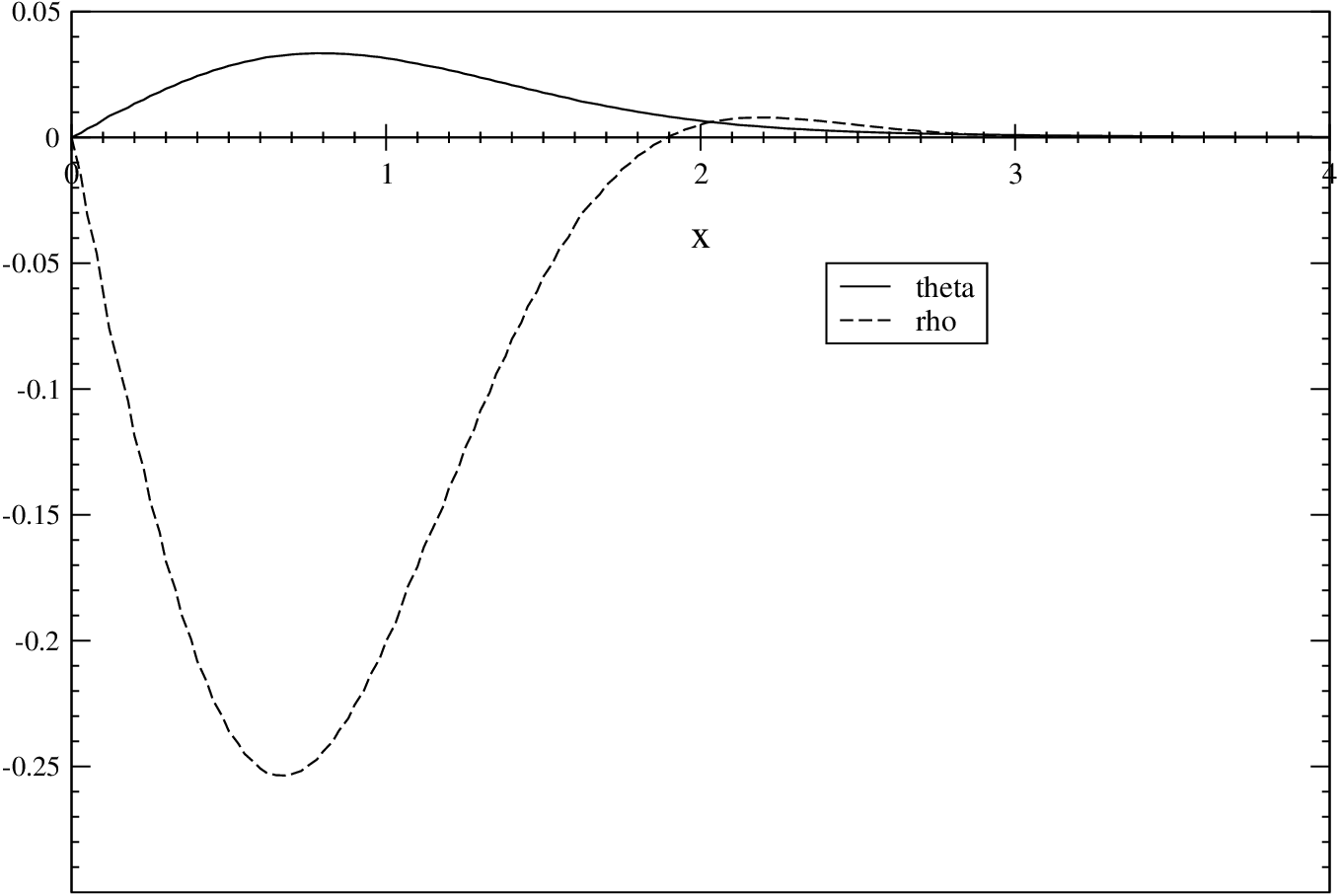}
\end{center}
\caption{The full line is the chiral angle $\protect\theta\left( x\right) $
and the dashed line is the source term $\protect\rho\left( x\right) $
defined in (\ref{rhox}).\ They are plotted as a function of $x=m_{\protect\pi%
}r$ for the $A=31$ configuration. The pion field $\protect\pi_{a}\left( \vec{%
r}\right) $ can be expressed in terms of $\protect\theta$ as $\protect\pi%
_{a}\left( \vec{r}\right) =f_{\protect\pi }\widehat{r}_{a}\protect\theta%
\left( r\right) $. In order to fit the figure, the source term $\protect\rho%
\left( x\right) $ is divided by $4\protect\pi m_{\protect\pi}f_{\protect\pi%
}^{2}$.}
\label{fig:rhotet}
\end{figure}

The configurations containing $A=13,15,29,31$ and $37$ nucleons in Table \ref
{table1} and $A=62,65$ and $67$ in Table \ref{table2} have appreciable
energy gaps.\ For these configurations, the chiral angle reaches maximum
values $0.03<\theta _{max}<0.04$.\ For the shape $\theta \left( x\right)
=axe^{-x}$ this corresponds to $a=e\theta _{max}$ in the vicinity of $\pm
0.1 $. In fact, the shape $axe^{-x}$ is quite similar to the calculated
shape of $\theta \left( x\right) $ as shown on Fig.\ref{fig:rhotet} and the
self-consistent single particle spectra are very similar to those which can
be read off Figs.\ref{fig:orb16} and \ref{fig:orbCa40}.

\begin{table}[h]
$
\begin{tabular}{|c|c|c|c|c|}
\hline
$\hbar\omega=12.71\;MeV$ &  &  &  &  \\ \hline
Closed $j$ shells & Filled $G$ shells & $A$ & Gap $\left( MeV\right) $ & $%
\theta_{max}$ \\ \hline
$
\begin{array}{c}
1s_{1/2}\;G=0,1 \\ 
1p_{3/2}\;G=1,2
\end{array}
$ & $\left( 1p_{1/2}\right) \;G=0$ & $13$ & $4.14$ & $0.038$ \\ \hline
$
\begin{array}{c}
1s_{1/2}\;G=0,1 \\ 
1p_{3/2}\;G=1,2
\end{array}
$ & $\left( 1p_{1/2}\right) \;G=1$ & $15$ & $4.27$ & $-0.031$ \\ \hline
$
\begin{array}{c}
1s_{1/2}\;G=0,1 \\ 
1p_{3/2}\;G=1,2 \\ 
1p_{1/2}\;G=0,1
\end{array}
$ & $\left( 1d_{5/2}\right) \;G=2$ & $21$ & $1.17$ & $-0.017$ \\ \hline
$
\begin{array}{c}
1s_{1/2}\;G=0,1 \\ 
1p_{3/2}\;G=1,2 \\ 
1p_{1/2}\;G=0,1
\end{array}
$ & $
\begin{array}{c}
\left( 1d_{5/2}\right) \;G=2 \\ 
\left( 2s_{1/2}\right) \;G=0
\end{array}
$ & $22$ & $1.4$ & $-0.04$ \\ \hline
$
\begin{array}{c}
1s_{1/2}\;G=0,1 \\ 
1p_{3/2}\;G=1,2 \\ 
1p_{1/2}\;G=0,1
\end{array}
$ & $\left( 1d_{5/2}\right) \;G=3$ & $23$ & $1.31$ & $0.02$ \\ \hline
$
\begin{array}{c}
1s_{1/2}\;G=0,1 \\ 
1p_{3/2}\;G=1,2 \\ 
1p_{1/2}\;G=0,1 \\ 
1d_{5/2}\;G=2,3
\end{array}
$ & $\left( 2s_{1/2}\right) \;G=0$ & $29$ & $2.51$ & $-0.035$ \\ \hline
$
\begin{array}{c}
1s_{1/2}\;G=0,1 \\ 
1p_{3/2}\;G=1,2 \\ 
1p_{1/2}\;G=0,1 \\ 
1d_{5/2}\;G=2,3
\end{array}
$ & $\left( 2s_{1/2}\right) \;G=1$ & $31$ & $2.92$ & $0.03$ \\ \hline
$\hbar\omega=10.27\;MeV$ &  &  &  &  \\ \hline
$
\begin{array}{c}
1s_{1/2}\;G=0,1 \\ 
1p_{3/2}\;G=1,2 \\ 
1p_{1/2}\;G=0,1 \\ 
1d_{5/2}\;G=2,3 \\ 
2s_{1/2}\;G=0,1
\end{array}
$ & $\left( 1d_{3/2}\right) \;G=1$ & $35$ & $1.64$ & $0.26$ \\ \hline
$
\begin{array}{c}
1s_{1/2}\;G=0,1 \\ 
1p_{3/2}\;G=1,2 \\ 
1p_{1/2}\;G=0,1 \\ 
1d_{5/2}\;G=2,3 \\ 
2s_{1/2}\;G=0,1
\end{array}
$ & $\left( 1d_{3/2}\right) \;G=2$ & $37$ & $2.55$ & $-0.023$ \\ \hline
\end{tabular}
$%
\caption{Closed $G$ shell configurations composed of $12<A<40$ nucleons.\
The first column lists the filled and closed $j$ shells which do not
contribute to the source of the pion field. The second column lists the
occupied filled $G$ shells. The column marked $A$ specifies the number of
nucleons in the configuration. The last two columns list the energy gap
separating filled and empty orbits and the maximum value of the chiral angle 
$\protect\theta.$}
\label{table1}
\end{table}

\begin{table}[ptb]
$
\begin{tabular}{|c|c|c|c|c|}
\hline
$\hbar\omega=10.27\;MeV$ &  &  &  &  \\ \hline
Closed $j$ shells & Closed $G$ shells & $A$ & Gap $\left( MeV\right) $ & $%
\theta_{max}$ \\ \hline
$1s,1p,2s,1d$ & $\left( 1f_{7/2}\right) \;G=3$ & $47$ & $0.56$ & $-0.011$ \\ 
\hline
$1s,1p,2s,1d$ & $\left( 1f_{7/2}\right) \;G=4$ & $49$ & $0.53$ & $0.013$ \\ 
\hline
$
\begin{array}{c}
1s,1p,2s,1d \\ 
1f_{7/2}\;G=3,4
\end{array}
$ & $\left( 2p_{3/2}\right) \;G=1$ & $59$ & $1.58$ & $-0.029$ \\ \hline
$
\begin{array}{c}
1s,1p,2s,1d \\ 
1f_{7/2}\;G=3,4
\end{array}
$ & $\left( 2p_{3/2}\right) \;G=2$ & $61$ & $1.15$ & $0.023$ \\ \hline
$
\begin{array}{c}
1s,1p,2s,1d \\ 
1f_{7/2}\;G=3,4
\end{array}
$ & $
\begin{tabular}{l}
$\left( 2p_{3/2}\right) \;G=2$ \\ 
$\left( 2p_{1/2}\right) \;G=0$%
\end{tabular}
$ & $62$ & $1.71$ & $0.043$ \\ \hline
$
\begin{array}{c}
1s,1p,2s,1d \\ 
1f_{7/2}\;G=3,4 \\ 
2p_{3/2}\;G=1,2
\end{array}
$ & $\left( 2p_{1/2}\right) \;G=0$ & $65$ & $2.18$ & $0.025$ \\ \hline
$
\begin{array}{c}
1s,1p,2s,1d \\ 
1f_{7/2}\;G=3,4 \\ 
2p_{3/2}\;G=1,2
\end{array}
$ & $\left( 2p_{1/2}\right) \;G=1$ & $67$ & $2.06$ & $-0.02$ \\ \hline
\end{tabular}
$%
\caption{The same as table \ref{table1} for $2p-1f$ shell nuclei.\ For all
these configurations the $1s,1p,2s,1d$ shells are filled.}
\label{table2}
\end{table}

\section{Discussion and conclusion.}

\setcounter{equation}{0}\renewcommand{\theequation}{\arabic{section}.%
\arabic{equation}}

In most nuclear structure calculations, the pi-nucleon interaction is
included in terms of pion exchange interactions between nucleons and the
possibility of generating a classical pion field is usually neglected. In
this calculation were have explored such a possibility by constructing
independent particle states which generate a hedgehog shaped classical pion
field. We have made a parameter free estimate of the generated pion field
and we found it to be small: the amplitude of the pion field, as measured by
the chiral angle, does not exceed $0.05$. We have found however, that such a
classical field can perturb significantly the spectrum of the nucleon
orbits. Such effects cannot be detected when Skyrme and Gogny interactions
are used or in most relativistic mean field calculations (see the reviews 
\cite{Heenen2003} and \cite{Vretenar2005}). Indeed, in such calculations,
the pi-nucleon interaction is absent and the effects of pion exchange
between nucleons is only included in terms of effective interactions (often
represented by scalar and vector fields) which simulate pion exchange in the
Fock term and 2-particle 2-hole excitations. The present calculation shows
that, in the mean field approximation, the pion-nucleon interaction does
give rise to a classical pion field in even-odd and odd-odd nuclei and it
points to possible deformations which could be taken into account in, for
example, generator coordinate calculations.

We have considered the simplest case of a hedgehog shaped pion field. We
have shown that such a field gives rise to nucleon orbits which have grand
spin $\vec{G}=\vec{J}+\vec{T}$ and that the hedgehog shape of the pion field
is a self-consistent symmetry in nuclei composed of closed $G$ shells (which
have $\vec{G}=\vec{J}+\vec{T}=0$). Closed $G$ shells are states with equal
average numbers of neutrons and protons. A single closed $G$ shell is
therefore an even-odd system of nucleons and two closed $G$ shells form an
odd-odd system. 

If such states were stable, they would generate $J=T$ rotational bands, as
they do for example in Skyrmions\cite{Witten83b} and chiral solitons (see
the review \cite{Ripka97}). However, because of the weak amplitude of the
pion field, the energy gaps created by the pion field in open $j$ shells are
not very large and it is unlikely that such rotational bands exist.\ A
rotating nucleus composed of closed $G$ shells would be seriously distorted
by Coriolis forces acting on the rotating intrinsic state\cite{Ripka88}.

The present calculation does not therefore allow a quantitative prediction
of the effect of the classical pion field on the spectra of some odd-even
and odd-odd nuclei.\ It is an exploratory calculation which indicates the
kind of deformations that are likely to be important in, for example,
generator coordinate calculations of the spectra of odd-even and odd-odd
nuclei. We have limited the present calculation to closed $G$ shells which
generate a hedgehog shaped pion field.\ There is no reason why this self
consistent symmetry should be favored in all nuclei.\ Other self-consistent
symmetries may be energetically favored when, for example, the neighboring
even-even nuclei have quadrupole deformations.

\appendix

\renewcommand{\theequation}{\Alph{section}.\arabic{equation}}

\setcounter{equation}{0}

\section{Some properties of grand spin states.}

Simple expressions may be derived for matrix elements between grand spin
states $\left| nljGM\right\rangle $ (\ref{ljgm}). If we adopt the
conventions (\ref{jgi}) and (\ref{jgizero}) to designate the four grand spin
states $\left| GMi\right\rangle $ which correspond to a given $\left(
G,M\right) $, then the matrix elements of $\vec{\sigma}\cdot\vec{\tau}$ are: 
\begin{equation*}
\left\langle GMi\left| \vec{\sigma}\cdot\vec{\tau}\right| G^{\prime
}M^{\prime}j\right\rangle
=\delta_{GG^{\prime}}\delta_{MM^{\prime}}M_{ij}^{\left( \sigma\tau\right) }
\end{equation*}
\begin{equation*}
M_{ij}^{\left( \sigma\tau\right) }=\left( 
\begin{array}{cccc}
1 & 0 & 0 & 0 \\ 
0 & 1 & 0 & 0 \\ 
0 & 0 & -\frac{2G+3}{2G+1} & \frac{4\sqrt{G\left( G+1\right) }}{2G+1} \\ 
0 & 0 & \frac{4\sqrt{G\left( G+1\right) }}{2G+1} & -\frac{2G-1}{2G+1}
\end{array}
\right) \allowbreak\allowbreak\allowbreak\;\;\left( G>0\right)
,\;\;\;M_{ij}^{\left( \sigma\tau\right) }=\left( 
\begin{array}{cc}
-3 & 0 \\ 
0 & 1
\end{array}
\right) \allowbreak\allowbreak\allowbreak\;\;\left( G=0\right)
\end{equation*}
The matrix elements of $\vec{\sigma}\cdot\widehat{r}$ are: 
\begin{equation*}
\left\langle GMi\left| \vec{\sigma}\cdot\widehat{r}\right| G^{\prime
}M^{\prime}j\right\rangle
=\delta_{GG^{\prime}}\delta_{MM^{\prime}}M_{ij}^{\left( r\sigma\right) }
\end{equation*}
\begin{equation*}
M_{ij}^{\left( r\sigma\right) }=\left( 
\begin{array}{cccc}
0 & 0 & 0 & -1 \\ 
0 & 0 & -1 & 0 \\ 
0 & -1 & 0 & 0 \\ 
-1 & 0 & 0 & 0
\end{array}
\right) \;\;\left( G>0\right) ,\;\;M_{ij}^{\left( r\sigma\right) }=\left( 
\begin{array}{cc}
0 & -1 \\ 
-1 & 0
\end{array}
\right) \;\;\left( G=0\right)
\end{equation*}
The matrix elements of $\left( \vec{\sigma}\cdot\widehat{r}\right) \left( 
\vec{\tau}\cdot\widehat{r}\right) $ are: 
\begin{equation*}
\left\langle GMi\left| \left( \vec{\sigma}\cdot\widehat{r}\right) \left( 
\vec{\tau}\cdot\widehat{r}\right) \right| G^{\prime}M^{\prime}j\right\rangle
=\delta_{GG^{\prime}}\delta_{MM^{\prime}}M_{ij}^{\left( r\sigma r\tau\right)
}
\end{equation*}
\begin{equation*}
M^{\left( r\sigma r\tau\right) }=\left( 
\begin{array}{cccc}
\frac{1}{2G+1} & 2\frac{\sqrt{\left( G\left( G+1\right) \right) }}{2G+1} & 0
& 0 \\ 
2\frac{\sqrt{\left( G\left( G+1\right) \right) }}{2G+1} & -\frac{1}{2G+1} & 0
& 0 \\ 
0 & 0 & -\frac{1}{2G+1} & \frac{2}{2G+1}\sqrt{G\left( G+1\right) } \\ 
0 & 0 & \frac{2}{2G+1}\sqrt{G\left( G+1\right) } & \frac{1}{2G+1}
\end{array}
\right) \;\;\;\left( G>0\right)
\end{equation*}
$\allowbreak$ 
\begin{equation*}
M^{\left( r\sigma r\tau\right) }=\allowbreak\left( 
\begin{array}{cc}
-1 & 0 \\ 
0 & -1
\end{array}
\right) \;\;\;\left( G=0\right)
\end{equation*}
The spin-orbit interaction $\vec{L}\cdot\vec{s}$ is diagonal: 
\begin{equation*}
\left\langle GMi\left| \vec{L}\cdot\vec{s}\right| G^{\prime}M^{\prime
}j\right\rangle
=\delta_{GG^{\prime}}\delta_{MM^{\prime}}\delta_{ij}A_{i}^{\left( ls\right) }
\end{equation*}
\begin{equation*}
A_{i}^{\left( ls\right) }=\left( \frac{1}{2}\left( G-1\right) ,-\frac
{1}{2}\left( G+2\right) ,\frac{1}{2}G,-\frac{1}{2}\left( G+1\right) \right)
\;\;\left( G>0\right)
\end{equation*}
\begin{equation*}
A_{i}^{\left( ls\right) }=\left( 0,-1\right) \;\;\left( G=0\right)
\end{equation*}
\qquad It is sometimes useful to use another grand spin basis $\left|
nlDGM\right\rangle $.\ It is defined by coupling the spin $\vec{s}$ and the
isospin $\vec{t}$ to a total spin $\vec{D}=\vec{s}+\vec{t}$ and by coupling
the orbital angular momenta $\vec{l}$ and $\vec{D}$ so as to form the grand
spin $\vec{G}=\vec{l}+\vec{D}$.\ For a given $\left( G,M\right) $ there are
four such states when $G>0$ and two states when $G=0$: 
\begin{equation*}
\begin{tabular}{|l|l|l|l|}
\hline
$\left| \left( D\right) GM1\right\rangle $ & $l=G-1$ & $D=1$ & $P=\left(
-1\right) ^{G+1}$ \\ \hline
$\left| \left( D\right) GM2\right\rangle $ & $l=G+1$ & $D=1$ & $P=\left(
-1\right) ^{G+1}$ \\ \hline
$\left| \left( D\right) GM3\right\rangle $ & $l=G$ & $D=0$ & $P=\left(
-1\right) ^{G}$ \\ \hline
$\left| \left( D\right) GM4\right\rangle $ & $l=G$ & $D=1$ & $P=\left(
-1\right) ^{G}$ \\ \hline
\end{tabular}
\;\;\;\;\left( G>0\right)
\end{equation*}
\begin{equation}
\begin{tabular}{llll}
$\left| \left( D\right) GM1\right\rangle $ & $l=0$ & $D=0$ & $P=+1$ \\ 
$\left| \left( D\right) GM2\right\rangle $ & $l=1$ & $D=1$ & $P=-1$%
\end{tabular}
\;\;\;\left( G=0\right)  \label{dgizero}
\end{equation}
The grand spin states $\left| \left( D\right) GMi\right\rangle $ are related
to the grand spin states $\left| GMi\right\rangle $ as follows: 
\begin{align*}
\left| GM1\right\rangle & =\left| \left( D\right) GM1\right\rangle \\
\left| GM2\right\rangle & =\left| \left( D\right) GM2\right\rangle \\
\left| GM3\right\rangle & =\sqrt{\frac{G+1}{2G+1}}\left| \left( D\right)
GM3\right\rangle +\sqrt{\frac{G}{2G+1}}\left| \left( D\right)
GM4\right\rangle \\
\left| ljGM4\right\rangle & =-\sqrt{\frac{G}{2G+1}}\left| \left( D\right)
GM3\right\rangle +\sqrt{\frac{G+1}{2G+1}}\left| \left( D\right)
GM4\right\rangle
\end{align*}

\begin{equation}
\left| G=0,i\right\rangle =\left| \left( D\right) G=0,i\right\rangle
\label{overl}
\end{equation}

\renewcommand{\theequation}{\Alph{section}.\arabic{equation}}

\setcounter{equation}{0}

\section{The contribution of the spin-orbit interaction to the mean field.}

Using the $v\left( 1,2\right) =v\left( 2,1\right) $ symmetry of the two body
spin-orbit interaction, its second quantized form (\ref{vls}) becomes the
sum of four terms: 
\begin{equation}
V_{ls}=V^{\left( 1\right) }+V^{\left( 2\right) }+V^{\left( 3\right)
}+V^{\left( 4\right) }
\end{equation}
which are: 
\begin{equation*}
V^{\left( 1\right) }=iW_{ls}\varepsilon _{ijk}\int d^{3}ra_{\alpha
}^{\dagger }a_{\beta }^{\dagger }\left\langle \alpha \left| \sigma
^{i}p^{j}\right| \vec{r}\right\rangle \left\langle \vec{r}\left|
p^{k}\right| \delta \right\rangle \left\langle \beta \left| \vec{r}\right.
\right\rangle \left\langle \vec{r}\left| \gamma \right. \right\rangle
a_{\gamma }a_{\delta }
\end{equation*}
\begin{equation*}
V^{\left( 2\right) }=iW_{ls}\varepsilon _{ijk}\int d1d2a_{\alpha }^{\dagger
}a_{\beta }^{\dagger }\left\langle \alpha \left| \sigma ^{i}\right| \vec{r}%
\right\rangle \left\langle \vec{r}\left| \delta \right. \right\rangle
\left\langle \beta \left| p^{j}\right| \vec{r}\right\rangle \left\langle 
\vec{r}\left| p^{k}\right| \gamma \right\rangle a_{\gamma }a_{\delta }
\end{equation*}
\begin{equation*}
V^{\left( 3\right) }=-iW_{ls}\varepsilon _{ijk}\int d1d2a_{\alpha }^{\dagger
}a_{\beta }^{\dagger }\left\langle \alpha \left| \sigma ^{i}p^{j}\right| 
\vec{r}\right\rangle \left\langle \vec{r}\left| \delta \right. \right\rangle
\left\langle \beta \left| \vec{r}\right. \right\rangle \left\langle \vec{r}%
\left| p^{k}\right| \gamma \right\rangle a_{\gamma }a_{\delta }
\end{equation*}
\begin{equation}
V^{\left( 4\right) }=-iW_{ls}\varepsilon _{ijk}\int d1d2a_{\alpha }^{\dagger
}a_{\beta }^{\dagger }\left\langle \alpha \left| \sigma ^{i}\right| \vec{r}%
\right\rangle \left\langle \vec{r}\left| p^{k}\right| \delta \right\rangle
\left\langle \beta \left| p^{j}\right| \vec{r}\right\rangle \left\langle 
\vec{r}\left| \gamma \right. \right\rangle a_{\gamma }a_{\delta }
\end{equation}
In these expressions, $\left| \alpha \right\rangle ,\left| \beta
\right\rangle ,...$ are nucleon single particle states expressed in an
arbitrary basis and a sum is assumed over repeated indices.

We evaluate the expectation value of $V_{ls}$ in the Slater determinant: 
\begin{equation}
\left| \Phi \right\rangle =a_{\lambda _{1}}^{\dagger }a_{\lambda
_{2}}^{\dagger }...a_{\lambda _{A}}^{\dagger }\left| 0\right\rangle \equiv
\prod_{\lambda \in F}a_{\lambda }^{\dagger }\left| 0\right\rangle 
\end{equation}
We use Wick's theorem to obtain: 
\begin{equation*}
\left\langle \Phi \left| V_{ls}\right| \Phi \right\rangle \equiv
\left\langle \Phi \left| V_{1}\right| \Phi \right\rangle +\left\langle \Phi
\left| V_{2}\right| \Phi \right\rangle +\left\langle \Phi \left|
V_{3}\right| \Phi \right\rangle +\left\langle \Phi \left| V_{4}\right| \Phi
\right\rangle 
\end{equation*}
\begin{equation*}
=2iW_{ls}\sum_{\lambda \mu \in F}\int d^{3}r\;\varepsilon _{ijk}\left\{
\left\langle \lambda \left| \sigma ^{i}p^{j}\right| \vec{r}\right\rangle
\left\langle \vec{r}\left| p^{k}\right| \lambda \right\rangle \left\langle
\mu \left| \vec{r}\right. \right\rangle \left\langle \vec{r}\left| \mu
\right. \right\rangle +\left\langle \lambda \left| \sigma ^{i}\right| \vec{r}%
\right\rangle \left\langle \vec{r}\left| \lambda \right. \right\rangle
\left\langle \mu \left| p^{j}\right| \vec{r}\right\rangle \left\langle \vec{r%
}\left| p^{k}\right| \mu \right\rangle \right. 
\end{equation*}
\begin{equation}
\left. -\left\langle \lambda \left| \sigma ^{i}p^{j}\right| \vec{r}%
\right\rangle \left\langle \vec{r}\left| \lambda \right. \right\rangle
\left\langle \mu \left| \vec{r}\right. \right\rangle \left\langle \vec{r}%
\left| p^{k}\right| \mu \right\rangle -\left\langle \lambda \left| \sigma
^{i}\right| \vec{r}\right\rangle \left\langle \vec{r}\left| p^{k}\right|
\lambda \right\rangle \left\langle \mu \left| p^{j}\right| \vec{r}%
\right\rangle \left\langle \vec{r}\left| \mu \right. \right\rangle \right\} 
\end{equation}
Using $p^{k}=\frac{1}{i}\nabla ^{k}$, integrating by parts and using the
antisymmetry of $\varepsilon _{ijk}$ we can reduce $\left\langle \Phi \left|
V_{ls}\right| \Phi \right\rangle $ to the form: 
\begin{equation*}
\left\langle \Phi \left| V_{ls}\right| \Phi \right\rangle
=2W_{ls}\sum_{\lambda \mu \in F}\int d^{3}r\varepsilon _{ijk}\left\langle
\lambda \left| \vec{r}\right. \right\rangle \left\langle \vec{r}\left|
\sigma ^{i}p^{k}\right| \lambda \right\rangle \left( \nabla ^{j}\left\langle
\mu \left| \vec{r}\right. \right\rangle \left\langle \vec{r}\left| \mu
\right. \right\rangle \right) 
\end{equation*}
\begin{equation*}
+2W_{ls}\sum_{\lambda \mu \in F}\int d^{3}r\varepsilon _{ijk}\left\langle
\mu \left| \vec{r}\right. \right\rangle \left\langle \vec{r}\left|
p^{k}\right| \mu \right\rangle \left( \nabla ^{j}\left\langle \lambda \left|
\sigma ^{i}\right| \vec{r}\right\rangle \left\langle \vec{r}\left| \lambda
\right. \right\rangle \right) 
\end{equation*}
\begin{equation*}
-2W_{ls}\sum_{\lambda \mu \in F}\int d^{3}r\varepsilon _{ijk}\left\langle
\lambda \left| \sigma ^{i}p^{j}\right| \vec{r}\right\rangle \left\langle 
\vec{r}\left| \lambda \right. \right\rangle \left\langle \mu \left| \vec{r}%
\right. \right\rangle \left\langle \vec{r}\left| ip^{k}\right| \mu
\right\rangle 
\end{equation*}
\begin{equation}
-2W_{ls}\sum_{\lambda \mu \in F}\int d^{3}r\varepsilon _{ijk}\left\langle
\lambda \left| \vec{r}\right. \right\rangle \left\langle \vec{r}\left|
\sigma ^{i}p^{k}\right| \lambda \right\rangle \left\langle \mu \left|
ip^{j}\right| \vec{r}\right\rangle \left\langle \vec{r}\left| \mu \right.
\right\rangle   \label{vexp1}
\end{equation}
This expression involves the densities: 
\begin{equation*}
n\left( \vec{r}\right) =\sum_{\lambda \in F}\left\langle \lambda \left| \vec{%
r}\right. \right\rangle \left\langle \vec{r}\left| \lambda \right.
\right\rangle \;\;\;\vec{p}\left( \vec{r}\right) =\sum_{\lambda \in
F}\left\langle \lambda \left| \vec{r}\right. \right\rangle \left\langle \vec{%
r}\left| \vec{p}\right| \lambda \right\rangle 
\end{equation*}
\begin{equation}
\vec{B}\left( \vec{r}\right) =\sum_{\lambda \in F}\left\langle \lambda
\left| \vec{r}\right. \right\rangle \left\langle \vec{r}\left| \left( \vec{%
\sigma}\times \vec{p}\right) \right| \lambda \right\rangle \;\;\;\vec{\sigma}%
\left( \vec{r}\right) =\sum_{\lambda \in F}\left\langle \lambda \left| \vec{r%
}\right. \right\rangle \left\langle \vec{r}\left| \vec{\sigma}\right|
\lambda \right\rangle 
\end{equation}
in terms of which: 
\begin{equation}
\left\langle \Phi \left| V_{ls}\right| \Phi \right\rangle =2W_{ls}\int d^{3}r
\left[ -\vec{B}\left( \vec{r}\right) \cdot \left( \vec{\nabla}n\left( \vec{r}%
\right) \right) -\vec{p}\left( \vec{r}\right) \cdot \left( \vec{\nabla}%
\times \vec{\sigma}\left( \vec{r}\right) \right) -i\vec{B}\left( \vec{r}%
\right) \cdot \vec{p}\left( \vec{r}\right) +i\vec{B}\cdot \vec{p}^{\ast
}\left( \vec{r}\right) \right]   \label{vexp2}
\end{equation}
Angular momentum coupling techniques can then readily be used to evaluate
the contribution of a closed $G$ shell to these densities, with the result
that the densities have the form: 
\begin{equation}
n\left( \vec{r}\right) =n\left( r\right) \;\;\;\vec{p}\left( \vec{r}\right) =%
\frac{1}{2i}\vec{\nabla}n\left( \vec{r}\right) =\frac{1}{2i}\widehat{r}\frac{%
dn}{dr}\;\;\;\vec{\sigma}\left( \vec{r}\right) =\widehat{r}\sigma \left(
r\right) \,\;\;\;\vec{B}\left( \vec{r}\right) =\widehat{r}\frac{1}{r}B\left(
r\right) 
\end{equation}
which could have been be guessed from the fact that closed $G$-shells are
invariant under rotations generated by spin + isospin. The second term of (%
\ref{vexp2}) therefore vanishes. The contribution of the spin orbit
interaction to the energy of a closed $G$-shell nucleus reduces thus to: 
\begin{equation}
\left\langle \Phi \left| V_{ls}\right| \Phi \right\rangle =-4W_{ls}\int
d^{3}r\;\vec{B}\left( \vec{r}\right) \cdot \left( \vec{\nabla}n\left( \vec{r}%
\right) \right) 
\end{equation}
which is the expression (\ref{vlsred}). The contribution can be expressed in
terms of the occupied orbits:
\begin{equation}
\left\langle \Phi \left| V_{ls}\right| \Phi \right\rangle
=-4W_{ls}\sum_{\lambda \mu \in F}\int d^{3}r\left\langle \lambda \left| \vec{%
r}\right. \right\rangle \left\langle \vec{r}\left| \left( \vec{\sigma}\times 
\vec{p}\right) _{i}\right| \lambda \right\rangle \left( \nabla
_{i}\left\langle \mu \left| \vec{r}\right. \right\rangle \left\langle \vec{r}%
\left| \mu \right. \right\rangle \right) 
\end{equation}
To obtain the contribution to the mean field, we calculate the (functional)
derivative:
\begin{equation}
\frac{\delta }{\delta \left\langle \lambda \left| \vec{r}s\right.
\right\rangle }\left\langle \Phi \left| V_{ls}\right| \Phi \right\rangle
=-4W_{ls}\left[ \left\langle \vec{r}s\left| \left( \vec{\sigma}\times \vec{p}%
\right) _{i}\right| \lambda \right\rangle \left( \nabla _{i}n\right)
-\left\langle \vec{r}s\left| \lambda \right. \right\rangle \left( \nabla
_{i}B_{i}\right) \right] 
\end{equation}
The contribution of the spin-orbit interaction to the mean field is thus the
operator:
\begin{equation}
h_{ls}=-4W_{ls}\left( \nabla _{i}n\right) \left( \vec{\sigma}\times \vec{p}%
\right) _{i}+4W_{ls}\left( \nabla _{i}B_{i}\right) 
\end{equation}
Since $n\left( \vec{r}\right) =n\left( r\right) $, we have:
\begin{equation*}
\left( \nabla _{i}n\right) \left( \vec{\sigma}\times \vec{p}\right) _{i}=%
\frac{1}{r}\frac{dn}{dr}\vec{r}\cdot \left( \vec{\sigma}\times \vec{p}%
\right) =-\frac{1}{r}\frac{dn}{dr}\vec{\sigma}\cdot \vec{L}
\end{equation*}
\begin{equation}
\nabla _{i}B_{i}=\nabla _{i}\widehat{r}_{i}B=\frac{1}{r}\frac{dB}{dr}+\frac{B%
}{r^{2}}
\end{equation}
so that:
\begin{equation}
h_{ls}=4W_{ls}\frac{1}{r}\frac{dn}{dr}\vec{\sigma}\cdot \vec{L}%
+4W_{ls}\left( \frac{1}{r}\frac{dB}{dr}+\frac{B}{r^{2}}\right) 
\end{equation}
which is the expression (\ref{hls1}).

\bibliographystyle{unsrt}
\bibliography{njl}

\end{document}